\documentclass[twocolumn,showpacs,aps,pra,amsfonts,amsmath,amssymb,superscriptaddress,floatfix]
{revtex4}
\usepackage[dvips]{graphicx}

\newcommand{\ket}[1]{\left | #1 \right \rangle}

\newcommand{\amp}[2]{\langle #1 | #2 \rangle}

\newcommand{\ex}[1]{\mathrm{e}^{#1}}

\newcommand{\eqr}[1]{Eq.~(\ref{#1})}

\newcommand{\affA}{%
     Quantum Information Technology Group, 
     National Institute of Information and Communications Technology
     (NICT), \\
     4-2-1 Nukui-kitamachi, Koganei, Tokyo 184-8795, Japan}
\newcommand{\affB}{%
     Department of Electronics and Electrical Engineering, 
     Keio University, \\
     3-14-1, Hiyoshi, Kohoku-ku, Yokohama, 223-8522, Japan}
\newcommand{\affC}{%
     CREST, Japan Science and Technology Agency, 
     1-9-9 Yaesu, Chuoh-ku, Tokyo 103-0028, Japan}
\newcommand{\affD}{%
     Institut f\"{u}r Optik, Information und Photonik,  
     Max-Planck Forschungsgruppe, Universit\"{a}t Erlangen-N\"{u}rnberg, 
     G\"{u}nther-Scharowsky str. 1, 91058, Erlangen, Germany}

\begin{document}
\title{Practical purification scheme for decohered coherent state superpositions via partial homodyne detection
}
%\date{\today}
%
\author{Shigenari Suzuki}
\address{\affA}%
\address{\affB}%
\author{Masahiro Takeoka}
\author{Masahide Sasaki}
\address{\affA}%
\address{\affC}%
\author{Ulrik L.~Andersen}
\address{\affD}%
\author{Fumihiko Kannari}
\address{\affB}%

\begin{abstract}

We present a simple protocol to purify 
%an a priori known superposition of coherent states, 
a coherent state superposition 
that has undergone a linear lossy channel. 
The scheme constitutes only a single beam splitter and a homodyne detector, 
and thus experimentally feasible. 
In practice, a superposition of coherent states is transformed into 
a classical mixture of coherent states by linear loss, 
which is usually the dominant decoherence mechanism in optical systems. 
We also address the possibility of producing a larger amplitude 
superposition state from decohered states, and show that in most cases 
the decoherence of the states are amplified along with the amplitude.

\end{abstract}
% 03.67.Hk Quantum communication  
% 42.50.-p Quantum optics

\maketitle

%%%%%%%%%%%%%%%%%%%%%%%%%%%%%%%%%%%%%%%%%%%%%%%%%%%%%%%%
\section{Introduction}
%%%%%%%%%%%%%%%%%%%%%%%%%%%%%%%%%%%%%%%%%%%%%%%%%%%%%%%%

In optical communication systems, 
information is often carried by states with Gaussian statistics;
the coherent state is a simple and very important example. 
The processing of these states in the quantum domain is now a major 
concern toward 
attaining ultimate capacity of bosonic channels 
\cite{Giovannetti04,Holevo98}, 
and also the construction of quantum networks and computers 
\cite{Yonezawa04,CV}. 
Toward these applications, it is important to generate and manipulate 
superpositions of pure Gaussian states. 
In fact, certain types of such superposition states 
can be used as ancillary states to build up 
a universal set of operations to construct an arbitrary processor
~\cite{LloydBraunstein99,Gottesman01,BartlettSanders02}. 
(See also ref~\cite{Sasaki04} for its experimental perspectives.) 
These states are essentially non-Gaussian states, 
and are easily destroyed even by a small amount of losses. 
In practice, it is not easy to prepare in labs the state 
that preserves enough purity and 
useful non-Gaussian properties. 
Therefore, one needs to develop the methods how to stabilize and/or purify 
such states.

A simple and more experimentally feasible example is 
a superposition of two coherent states with small amplitudes. 
Such a state can be conditionally generated from a squeezed vacuum 
(Gaussian state) by using the photon number measurement~\cite{dakna97}. 
Recently, a proof-of-principle experiment of this idea was successfully 
presented in ref~\cite{wenger04}, where an on/off-type detector 
(which clicks upon the arrival of a photon) 
was used rather than a photon number resolving detector. 
There is, however, still a distance between the observed state and 
the ideal one due to various losses.

In this paper, we develop a feasible scheme to purify 
such coherent superposition states (CSSs) 
that has undergone linear dissipation. 
We consider a specific scenario of the purification, 
but it meets the current experimental situations and 
is based only on the use of a single beam splitter 
and a single homodyne detector, 
which probabilistically selects out favourable events.
The scenario we treat is the following; 
Consider a CSS of an equal superposition of 
two distinct coherent states, 
$|\alpha\rangle + e^{i\varphi} |-\alpha\rangle$, 
where the amplitude $\alpha$ and the phase $\varphi$ are 
a priori known and $\alpha$ is relatively small. 
This state decoheres into a mixed state by the linear loss and represented by  
a classical mixture of a CSS and two coherent states.
Our task is then to amplify the probability 
of the CSS as efficiently as possible.

The CSS itself has also been discussed as a resource of 
various optical quantum information processing protocols
\cite{vanenk01, jeong01, jeong02-pra, ralph02, ralph03, jeong02-qic,
clausen02, cochrane99}.  
As a related topic to our study, 
Glancy {\it et al.} have proposed a method to restore a CSS, 
whose weights and phase information of the superposition are unknown, 
by applying a three-bit quantum error correcting code \cite{glancy04}. 
However, this proposal requires almost perfect photon number resolving 
detectors and pure CSSs as ancillary states. 
In our study we make no use of non-Gaussian ancillary states 
in the purifaction method.

\begin{figure}[h]
\begin{center}
 \includegraphics[width=0.9\linewidth]{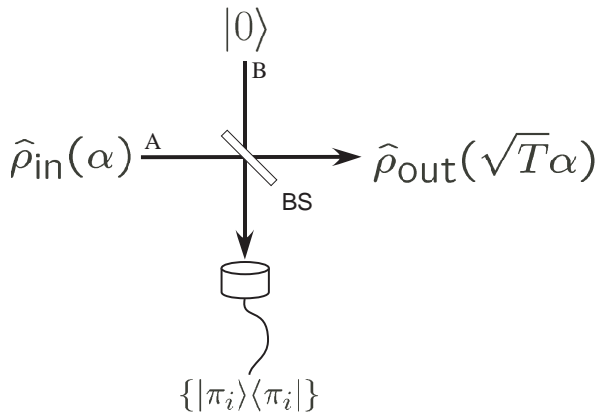}
 \caption{Schematic of the conditional purification scheme using partial measurement. BS: Beam splitter with transmittivity $T$.}
 \label{fig:scheme}
\end{center}
\end{figure}

The schematic of our proposal is shown in Fig.~\ref{fig:scheme}.
A fraction of the input state is tapped off
by a beam splitter and subsequently measured with a homodyne detector. 
From the measurement outcomes we
try to distinguish the CSS from the mixture of coherent states through 
the differences of their 
wave functions. Based on the success of this estimation, 
the coherence of the CSS is retrieved 
through selection of favourable events. 
We show that homodyne measurement is one of the optimal measurements 
for the most efficient purification given such a simple setup.

Using the purification protocol in Fig.~\ref{fig:scheme}, 
the amplitude of the coherent state components in the CCS is 
inevitably degraded due to 
the conditional selection of events. 
So we then address the possibility to recover the amplitude 
by using the CSS amplification scheme proposed in refs \cite{lund04,jeong05}, 
where it has been shown that a CSS with small amplitudes of its constituents 
is equivalently generated by squeezing a single photon state and 
the amplitudes can be amplified using linear optics and on/off detectors.

This paper is organized as follows. 
In section \ref{sec:2}, the influence of linear dissipation to the CSS is briefly reviewed.  
In section \ref{sec:3}, we discuss the purification scheme in general terms without specifying the exact measurement strategy applied to the tapped off state.
In section \ref{sec:4}, we consider the example where a homodyne detector is used and show that it is one of the optimal measurement strategies.
In section \ref{sec:5}, we discuss the amplification of CSS using two
CSSs that has suffered from linear loss, and finally in section \ref{sec:6} we conclude.

%%%%%%%%%%%%%%%%%%%%%%%%%%%%%%%%%%%%%%%%%%%%%%%%%%%%%%%%
\section{Decoherence of CSS in a lossy quantum
 channel\label{sec:2}}
%%%%%%%%%%%%%%%%%%%%%%%%%%%%%%%%%%%%%%%%%%%%%%%%%%%%%%%%
In this section we briefly summarize the evolution of CSS's under linear dissipation. 
Let us first define a superposition of two coherent states (CSS) 
with the amplitudes $\ket{\pm\beta}$ and the relative phase $\varphi$ as 
%%%%%%%%%%%%%%%%%%%%%%%%%%%%%%%%%%%%%%%%%%%%%%%%%%%%%%%%
\begin{equation}
\label{eq:cat}
|\psi_{\varphi}(\beta)\rangle 
= \frac{1}{\sqrt{N_{\varphi}(\beta)}}
\left( |\beta\rangle + e^{i\varphi} |-\beta\rangle \right),
\end{equation}
%%%%%%%%%%%%%%%%%%%%%%%%%%%%%%%%%%%%%%%%%%%%%%%%%%%%%%%%
where
%%%%%%%%%%%%%%%%%%%%%%%%%%%%%%%%%%%%%%%%%%%%%%%%%%%%%%%%
\begin{equation}
\label{eq:norm}
N_{\varphi}(\beta)=2 ( 1+\cos\varphi~ e^{-2\beta^2} ). 
\end{equation}
%%%%%%%%%%%%%%%%%%%%%%%%%%%%%%%%%%%%%%%%%%%%%%%%%%%%%%%%
We assume $\beta$ to be real without loss of generality.

In practice, the most dominant loss in 
optical channels is due to a linear interaction 
between the signal mode and the vacuum environment. 
This is modeled by coupling the signal mode to the environment, letting them evolve for some time and finally trace out the states of the environment.
A completely positive (CP) map describing such a linear interaction between the signal and the environment is 
given by \cite{QuantumOptics}
%%%%%%%%%%%%%%%%%%%%%%%%%%%%%%%%%%%%%%%%%%%%%%%%%%%%%%%%
\begin{eqnarray}
\label{eq1.1}
{\cal L}_L & = &
\exp \left[ -\ln \eta \left( {\cal K}_- - {\cal K}_0 \right) \right], 
\end{eqnarray}
%%%%%%%%%%%%%%%%%%%%%%%%%%%%%%%%%%%%%%%%%%%%%%%%%%%%%%%%
where ${\cal K}_-$ and ${\cal K}_0$ are the superoperators defined as
%%%%%%%%%%%%%%%%%%%%%%%%%%%%%%%%%%%%%%%%%%%%%%%%%%%%%%%%
\begin{eqnarray}
\label{eq1.2}
{\cal K}_- \hat{X} & = & \hat{a} \hat{X} \hat{a}^{\dagger} \qquad
{\cal K}_0 \hat{X} = {\textstyle\frac{1}{2}} 
  ( \hat{a} \, \hat{a}^{\dagger} \hat{X} + \hat{X} \hat{a}^{\dagger}
  \hat{a} ) ,
\end{eqnarray}
%%%%%%%%%%%%%%%%%%%%%%%%%%%%%%%%%%%%%%%%%%%%%%%%%%%%%%%%
for an arbitrary operator $\hat{X}$. The transmission of the environment is denoted by $\eta$ and thus the linear loss is given by $1-\eta$.
Using the transformation 
%%%%%%%%%%%%%%%%%%%%%%%%%%%%%%%%%%%%%%%%%%%%%%%%%%%%%%%%
\begin{eqnarray}
\label{eq1.3a}
{\cal L}_L | \alpha_1 \rangle \langle \alpha_2 | 
  & = & \exp 
  \left[ - \frac{1}{2} (1-\eta) 
   \left(
    | \alpha_1 |^2 + | \alpha_2 |^2 - 2 \alpha_1 \alpha_2^* 
   \right) 
  \right] 
  \nonumber\\ & & 
  \times| \alpha_1 \sqrt{\eta} \rangle \langle \alpha_2 \sqrt{\eta} | .
\end{eqnarray}
%%%%%%%%%%%%%%%%%%%%%%%%%%%%%%%%%%%%%%%%%%%%%%%%%%%%%%%%
we find that the CSS, $\hat{\rho}_{C_{\varphi}}(\beta)=
|\psi_{\varphi}(\beta)\rangle\langle\psi_{\varphi}(\beta)|$, under linear dissipation transforms into the mixture 
%%%%%%%%%%%%%%%%%%%%%%%%%%%%%%%%%%%%%%%%%%%%%%%%%%%%%%%%
\begin{equation}
\label{eq:decohered_cat}
{\cal L}_L \hat{\rho}_{C_{\varphi}}(\beta)
= p \hat{\rho}_{C_{\varphi}}(\sqrt{\eta}\beta) 
+ (1-p) \hat{\rho}_0(\sqrt{\eta}\beta),
\end{equation}
%%%%%%%%%%%%%%%%%%%%%%%%%%%%%%%%%%%%%%%%%%%%%%%%%%%%%%%%
where $\hat{\rho}_0(\sqrt{\eta}\beta)$ is a mixed state of two 
coherent states $\hat{\rho}_0(\sqrt{\eta}\beta)
=\frac{1}{2}(|\sqrt{\eta}\beta\rangle\langle\sqrt{\eta}\beta| 
+ |-\sqrt{\eta}\beta\rangle\langle-\sqrt{\eta}\beta|)$. 
Here the fraction of the original state in this mixed state is given by 
%%%%%%%%%%%%%%%%%%%%%%%%%%%%%%%%%%%%%%%%%%%%%%%%%%%%%%%%
\begin{equation}
\label{eq:p}
p = \frac{1 + \cos\varphi \, e^{-2\eta\beta^2}}{
1 + \cos\varphi \, e^{-2\beta^2}} e^{-2(1-\eta)\beta^2}.
\end{equation}
%%%%%%%%%%%%%%%%%%%%%%%%%%%%%%%%%%%%%%%%%%%%%%%%%%%%%%%%
As expected, the linear lossy interaction with the environment transforms the original pure CSS state into a classical mixture. The aim of this paper is to develop a method by which the decohered state can be probabilistically purified to produce a smaller number of states with higher purity, or in other words; the aim is to increase the value of $p$.

Finally, we note that the purity defined 
by ${\rm Tr}[\hat{\rho}^2]$ is not a suitable figure of merit 
for our purpose. 
One can easily show that, for the mixed CSS in Eq.~(\ref{eq:decohered_cat}), 
a smaller amplitude gives a higher purity of the state 
even for small $p$. 
This is because such a state is close to a vacuum state 
which is completely pure. 
However, it is obvious that it does not mean the state is close to a pure CSS.

%%%%%%%%%%%%%%%%%%%%%%%%%%%%%%%%%%%%%%%%%%%%%%%%%%%%%%%%%%%%%%
\section{Purification via partial measurement\label{sec:3}}
%%%%%%%%%%%%%%%%%%%%%%%%%%%%%%%%%%%%%%%%%%%%%%%%%%%%%%%%%%%%%%
Our proposed scheme for probabilistic purification of CSSs is illustrated in Fig.~\ref{fig:scheme}. The state under interrogation is mixed with the vacuum at a beam splitter with the transmission coefficient $T$. The reflected beam is subsequently measured using a positive operator-valued measure (POVM) generally described by a set of positive operators $\{|\pi_i\rangle\langle\pi_i| \}$ with $\sum_i |\pi_i\rangle\langle\pi_i|=\hat{1}$. Depending on the outcome of this measurement, the transmitted state is probabilistically purified: If the outcomes fall within a certain detection range, the transmitted state is purified, otherwise the state turns out to be worse and is discarded.

Let us first describe the function of the purification scheme when an already pure CSS 
$\hat\rho_{C_\varphi} (\alpha)=|\psi_\varphi (\alpha) \rangle\langle\psi_\varphi (\alpha)|$ is injected into it.  
In this case, the output of the beam splitter (described by the unitary operator $\hat{B}(T)$) is given by the entangled state
%%%%%%%%%%%%%%%%%%%%%%%%%%%%%%%%%%%%%%%%%%%%%%%%%%%%%%%%
\begin{eqnarray}
\label{eq:BS}
\hat{B}(T) |\psi_{\varphi}(\alpha)\rangle_A |0\rangle_B & = & 
\frac{1}{\sqrt{N_{\varphi}(\alpha)}} \left(
|\sqrt{T}\alpha\rangle_A|\sqrt{R}\alpha\rangle_B 
\right. \nonumber\\ & & \left.
+ e^{i\varphi} 
|-\sqrt{T}\alpha\rangle_A|-\sqrt{R}\alpha\rangle_B 
\right),
\nonumber\\
\end{eqnarray}
%%%%%%%%%%%%%%%%%%%%%%%%%%%%%%%%%%%%%%%%%%%%%%%%%%%%%%%%
where $R=1-T$ is the reflectivity of the
beam splitter. If we now consider only one output and trace out the other, the pure CSS is transformed into a mixed state as a result of the coupling to the vacuum environment. 
The purification protocol now works by making a measurement on the ``environmental'' mode (mode B) and selecting the outcomes that map $|\psi_\varphi(\alpha) \rangle$ back onto a pure CSS at the output. 
It implies that the measurement outcomes for successful events 
have to satisfy
%%%%%%%%%%%%%%%%%%%%%%%%%%%%%%%%%%%%%%%%%%%%%%%%%%%%%%%%
\begin{equation}
\label{eq:condition}
\langle\pi_j^s|-\sqrt{R}\alpha\rangle = 
e^{i\theta_j}\langle\pi_j^s|\sqrt{R}\alpha\rangle, 
\end{equation}
%%%%%%%%%%%%%%%%%%%%%%%%%%%%%%%%%%%%%%%%%%%%%%%%%%%%%%%%
where the superscript $s$ denotes the successful events. 
The unnormalized conditional output for such outcomes is given by 
%%%%%%%%%%%%%%%%%%%%%%%%%%%%%%%%%%%%%%%%%%%%%%%%%%%%%%%%
\begin{eqnarray}
\label{eq:cond_output}
& & _B\langle\pi_j^s|\hat{B}(T) 
|\psi_{\varphi}\rangle_A |0\rangle_B 
\nonumber\\ & & 
= \frac{1}{N_{\varphi}(\alpha)^{1/2}}
\langle\pi_j^s|\sqrt{R}\alpha\rangle  \left(
|\sqrt{T}\alpha\rangle + e^{i(\varphi+\theta_j)} |-\sqrt{T}\alpha\rangle
\right) 
\nonumber\\ & & 
= \langle\pi_j^s|\sqrt{R}\alpha\rangle \left( 
\frac{N_{\varphi+\theta_j} (\sqrt{T}\alpha)}{
N_\varphi (\alpha)} \right)^{1/2}
|\psi_{\varphi+\theta_j} (\sqrt{T}\alpha) \rangle,
\end{eqnarray}
%%%%%%%%%%%%%%%%%%%%%%%%%%%%%%%%%%%%%%%%%%%%%%%%%%%%%%%%
The output state of the purifier is therefore a pure CSS with a reduced amplitude and a phase shift with respect to the input state. This means that by applying measurements that fulfill \eqr{eq:condition}, it is possible to recover the purity of the input CSS.

Let us now consider the purification of a decohered CSS. The input to the purifier is now the mixture 
%%%%%%%%%%%%%%%%%%%%%%%%%%%%%%%%%%%%%%%%%%%%%%%%%%%%%%%%
\begin{equation}
\label{eq:initial_state}
\hat{\rho}_{\rm in} (\alpha) = 
p_{\rm in} \hat{\rho}_{C_{\varphi}}(\alpha) 
+ (1-p_{\rm in}) \hat{\rho}_0(\alpha). 
\end{equation}
%%%%%%%%%%%%%%%%%%%%%%%%%%%%%%%%%%%%%%%%%%%%%%%%%%%%%%%%
The reflected part of the state is measured using the above mentioned POVM and events corresponding to condition (\ref{eq:condition}) are selected. The resulting density operator of the output state conditioned on this measurement is given by
%%%%%%%%%%%%%%%%%%%%%%%%%%%%%%%%%%%%%%%%%%%%%%%%%%%%%%%%
\begin{eqnarray}
 \label{eq:purified_state}
  \hat{\rho}_{\rm out} (\sqrt{T}\alpha)
  & = & \frac{
  p_{\rm in} P_{C_{\varphi}} \hat{\rho}_{C_{\varphi+\theta_j}}
  (\sqrt{T}\alpha) + 
  (1-p_{\rm in}) P_0 \hat{\rho}_0 (\sqrt{T}\alpha)}{
  p_{\rm in} P_{C_{\varphi}} + (1-p_{\rm in}) P_0 } \nonumber\\
 & \equiv & 
  p_{\rm out} \hat{\rho}_{C_{\varphi+\theta_j}} (\sqrt{T}\alpha) + 
  (1-p_{\rm out}) \hat{\rho}_0 (\sqrt{T}\alpha)
\end{eqnarray}
%%%%%%%%%%%%%%%%%%%%%%%%%%%%%%%%%%%%%%%%%%%%%%%%%%%%%%%%
where 
%%%%%%%%%%%%%%%%%%%%%%%%%%%%%%%%%%%%%%%%%%%%%%%%%%%%%%%%
\begin{eqnarray}
P_{C_{\varphi}} & = & 
{\rm Tr}_A \left[ _B \langle\pi_j^s| \hat{B}(T) \left(
\hat{\rho}_{C_{\varphi}} (\alpha) \otimes |0\rangle\langle0| 
\right)
\hat{B}^{\dagger}(T) |\pi_j^s\rangle_B \right] 
\nonumber\\ 
& = & \left|\langle\pi_j^s|\sqrt{R}\alpha\rangle\right|^2 
\frac{1+\cos(\varphi+\theta_j) \, e^{-2T\alpha^2}}{
1+\cos\varphi \, e^{-2\alpha^2}} 
\end{eqnarray}
and
\begin{eqnarray}
\label{eq:det_probability_c}
P_0 & = & {\rm Tr}_A \left[ _B \langle\pi_j^s| \hat{B}(T) \left(
\hat{\rho}_0 (\alpha) \otimes |0\rangle\langle0| 
\right)
\hat{B}^{\dagger}(T) |\pi_j^s\rangle_B \right] 
\nonumber\\
& = & \left|\langle\pi_j^s|\sqrt{R}\alpha\rangle\right|^2  
\label{eq:det_probability_0}
\end{eqnarray}
%%%%%%%%%%%%%%%%%%%%%%%%%%%%%%%%%%%%%%%%%%%%%%%%%%%%%%%%
correspond to the probability distributions for detecting the outcomes $\pi_j^s$ in the states $\hat{\rho}_{C_{\varphi}} (\alpha)$ and 
$\hat{\rho}_0 (\alpha)$, respectively. 
As a consequence, one finds the fraction of the CSS
after the purification as 
%%%%%%%%%%%%%%%%%%%%%%%%%%%%%%%%%%%%%%%%%%%%%%%%%%%%%%%%
\begin{eqnarray}
 \label{eq:p_out}
  p_{\rm out} 
  & = & \frac{P_{C_{\varphi}}}{
  p_{\rm in} P_{C_{\varphi}} + (1-p_{\rm in}) P_0} p_{\rm in}
  \nonumber\\ 
& = & \frac{1}{p_{\rm in} + P_0/P_{C_{\varphi}} (1-p_{\rm in})} p_{\rm
 in} .  
\end{eqnarray}
%%%%%%%%%%%%%%%%%%%%%%%%%%%%%%%%%%%%%%%%%%%%%%%%%%%%%%%%
The condition for successful purification 
$p_{\rm out} > p_{\rm in}$ is then given by
%%%%%%%%%%%%%%%%%%%%%%%%%%%%%%%%%%%%%%%%%%%%%%%%%%%%%%%%
\begin{equation}
\label{eq:purif_codition}
\frac{P_0}{P_{C_{\varphi}}} = \frac{1+\cos\varphi \, e^{-2\alpha^2}}{
1+\cos(\varphi+\theta_j) \, e^{-2T\alpha^2}} < 1, 
\end{equation}
%%%%%%%%%%%%%%%%%%%%%%%%%%%%%%%%%%%%%%%%%%%%%%%%%%%%%%%%
The maximum purification efficiency is achieved when 
$P_0/P_{C_{\varphi}}$ is minimized. There are two free parameters controlling the magnitude of this fraction, namely $\theta_j$ and $T$. First we note that the phase $\theta_j$ is optimized, that is $P_0/P_{C_{\varphi}}$ is minimized when $\theta_j=-\varphi$.  
Second, it is clear that the fraction (\ref{eq:purif_codition}) decreases with decreasing transmissions $T$. This, on the other hand, means that the amplitude of the CSS is decreasing. Therefore, there is a trade-off between the efficiency of the purification protocol and the amplitude of the purified CSS. Since large amplitude CSSs 
are desired for many applications, we consider the possibility of amplifying the degraded CSS in Sec.~V.

The optimal POVM for the purification protocol is therefore 
the one that obeys \eqr{eq:condition} and simultaneously minimizes the fraction (\ref{eq:purif_codition}) by satisfying the constraint $\theta_j+\varphi=0$. These conditions provide a nice flexibility to chose among different measurement strategies. Two well known measurement strategies that satisfy the requirements are photon number resolving detectors and homodyne detectors. If however the photon number resolving detector is used to execute the purification protocol, successful demonstration will only be obtained if the phase between the coherent state components in the CSS, is $\varphi=0$ or $\pi$. Furthermore, these detectors are experimentally very challenging to construct and thus, to date, they are not an integral part of the standard lab.

Fortunately, homodyne detectors (which have matured much further than photon number resolving detectors) also meet the requirements for successful purification. In fact by using homodyne detectors there are no restrictions on the phase $\varphi$.
In the following section we investigate in greater details the purification of CSS using homodyne detection.

%%%%%%%%%%%%%%%%%%%%%%%%%%%%%%%%%%%%%%%%%%%%%%%%%%%%%%%%%%%%%%%
\section{Purification via homodyne measurement\label{sec:4}}
%%%%%%%%%%%%%%%%%%%%%%%%%%%%%%%%%%%%%%%%%%%%%%%%%%%%%%%%%%%%%%%
A homodyne detector measures a quadrature of the electromagnetic field, with the specific quadratures being determined by the phase, $\lambda$, of the intrinsic local oscillator. It is mathematically described by an operator $|x_\lambda\rangle\langle x_\lambda|$ which projects the state onto a continuous set of quadrature eigenstates $\{\ket{x_\lambda} \}$ 
with $\lambda$ denoting the quadrature phase. Specific and important examples are the position and momentum eigenstates corresponding to $\lambda=0$ and $\lambda=\pi/2$, respectively.
The probability amplitude associated with the detection of a quadrature eigenvalue, $x_\lambda$, from the coherent state $\ket{\beta}$ by homodyne detection is given by 
%%%%%%%%%%%%%%%%%%%%%%%%%%%%%%%%%%%%%%%%%%%%%%%%%%%%%%%%
\begin{eqnarray}
\langle x_\lambda | \beta \rangle
&=\pi^{-1/4} \exp &\left[
	-\frac{1}{2} x_\lambda^2 -\sqrt{2} e^{-i\lambda} x_\lambda \beta
	\right.
	\nonumber\\
 &&	\left. 
	 -\frac{1}{2} e^{-2i\lambda} \beta^2 -\frac{1}{2}|\beta|^2
      \right].
\end{eqnarray}
%%%%%%%%%%%%%%%%%%%%%%%%%%%%%%%%%%%%%%%%%%%%%%%%%%%%%%%%
The necessary condition for our purification scheme to work as quoted  
in \eqr{eq:condition} is fulfilled when $\lambda=\pi/2$, that is when the momentum eigenstate is measured. In that case we obtain 
%%%%%%%%%%%%%%%%%%%%%%%%%%%%%%%%%%%%%%%%%%%%%%%%%%%%%%%%
\begin{eqnarray}
\amp{x_{\pi/2}}{-\beta}
=e^{i 2\sqrt{2} x_{\pi/2} \beta} \amp{x_{\pi/2}}{\beta},
\label{eq:condition-hd}
\end{eqnarray}
%%%%%%%%%%%%%%%%%%%%%%%%%%%%%%%%%%%%%%%%%%%%%%%%%%%%%%%%
from which we see that the phase shift introduced by the measurement depends on $x_{\pi/2}$ as 
%%%%%%%%%%%%%%%%%%%%%%%%%%%%%%%%%%%%%%%%%%%%%%%%%%%%%%%%
\begin{eqnarray}
\theta=2\sqrt{2} x_{\pi/2} \beta.
\label{eq:theta_hd}
\end{eqnarray}
%%%%%%%%%%%%%%%%%%%%%%%%%%%%%%%%%%%%%%%%%%%%%%%%%%%%%%%%
Note that the homodyne measurement always allows us to 
choose the appropriate $x_{\pi/2}$ to optimize $\theta$ for any given $\varphi$.

In the following we set $k \equiv x_{\pi/2}$. 
The unnormalized conditional output corresponding to \eqr{eq:cond_output} 
is then given by 
%%%%%%%%%%%%%%%%%%%%%%%%%%%%%%%%%%%%%%%%%%%%%%%%%%%%%%%%
\begin{eqnarray}
\lefteqn{_B\langle k |\hat{B}(T) | \psi_{\varphi}\rangle_A |0\rangle_B} 
\nonumber\\
&=&
 \frac{1}{N_{\varphi}(\alpha)^{1/2}}
 \amp{k}{\sqrt{R}\alpha}
 \left(
  |\sqrt{T}\alpha\rangle + e^{i(\varphi+\theta)} |-\sqrt{T}\alpha\rangle
 \right)  \nonumber\\
 &=&
  \frac{\ex{-\frac{1}{2}k^2 -i \sqrt{2R}\alpha k}}{\pi^{1/4}}
  \sqrt{\frac{1 + \cos(\varphi +2\sqrt{2R}\alpha k) \ex{-2T\alpha^2}
  }{1 +\cos \varphi ~\ex{-2\alpha^2}}}
  \nonumber\\
&& \times \ket{\psi_{\varphi+2\sqrt{2R}\alpha k}(\sqrt{T}\alpha)}_A,
  \label{eq:cond_output_hd}
\end{eqnarray}
%%%%%%%%%%%%%%%%%%%%%%%%%%%%%%%%%%%%%%%%%%%%%%%%%%%%%%%%
and one obtains the probability distributions 
for detecting the $\hat{\rho}_{C_\varphi}(\alpha)$ and $\hat{\rho}_0(\alpha)$ as

%%%%%%%%%%%%%%%%%%%%%%%%%%%%%%%%%%%%%%%%%%%%%%%%%%%%%%%%
\begin{eqnarray}
P_{C_{\varphi}} (k)
&=& \frac{\ex{-k^2}}{\pi^{1/2}}
 \frac{1 + \cos(\varphi +2\sqrt{2R}\alpha k) \ex{-2T\alpha^2}}{1 +\cos
 \varphi ~\ex{-2\alpha^2}}
\nonumber\\
\label{eq:det_probability_c_hd}
\end{eqnarray}
and 
\begin{eqnarray}
P_0 (k) 
 &=& \frac{\ex{-k^2}}{\pi^{1/2}}, 
\label{eq:det_probability_0_hd}
\end{eqnarray}
%%%%%%%%%%%%%%%%%%%%%%%%%%%%%%%%%%%%%%%%%%%%%%%%%%%%%%%%
respectively.

\begin{figure}[h]
 \begin{center}
  \includegraphics[width=0.9\linewidth]{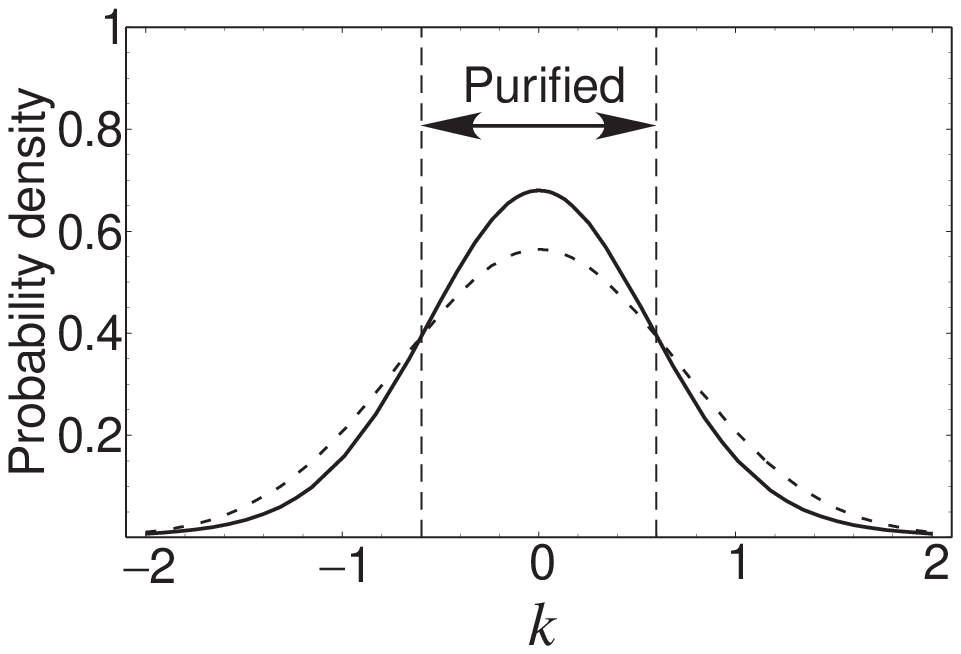}
  \caption{Probability quadrature distributions of a CSS $P_{C_{0}}$ with $\varphi =0$ (solid line) and a mixture of coherent states 
  $P_0$ (dotted line). $T=0.5$ and $\alpha=1$. The purification protocol is successful when the measurement outcome falls within the interval indicated by the dashed vertical lines.}
  \label{fig:cond_k_p}
 \end{center}
\end{figure}

\begin{figure}[h]
 \begin{center}
  \includegraphics[width=0.9\linewidth]{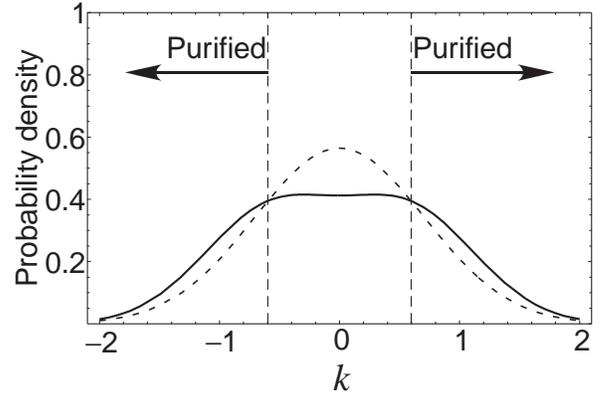}
  \caption{Probability quadrature distributions for 
  a CSS $P_{C_{\varphi}}$ with $\varphi =\pi$(solid line) and a mixture of coherent states 
  $P_0$ (dotted line). $T=0.5$ and $\alpha=1$. The purification protocol is successful when the measurement outcome falls outside the interval indicated by the dashed vertical lines.}
  \label{fig:cond_k_m}
 \end{center}
\end{figure}

\begin{figure}[h]
 \begin{center}
  \includegraphics[width=0.9\linewidth]{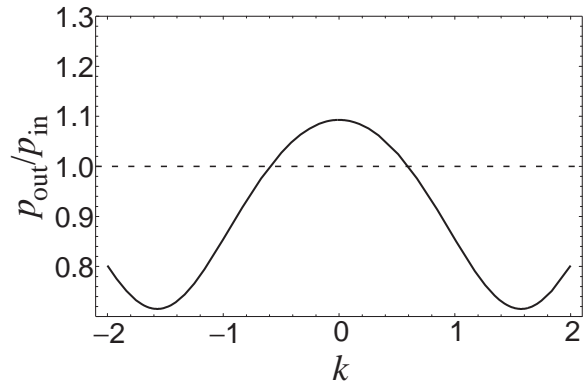}
  \caption{Purification efficiency for 
  the CSS with $\varphi=0$ conditioned on 
  the homodyne detection outcomes. The phase of the local oscillator 
  is chosen to measure the quadrature $k \equiv x_{\pi/2}$. 
  $T=0.5$, $\alpha=1$, and $p_{\rm in}=0.5$.} 
  \label{gainp}
 \end{center}
\end{figure}

\begin{figure}[h]
 \begin{center}
  \includegraphics[width=0.9\linewidth]{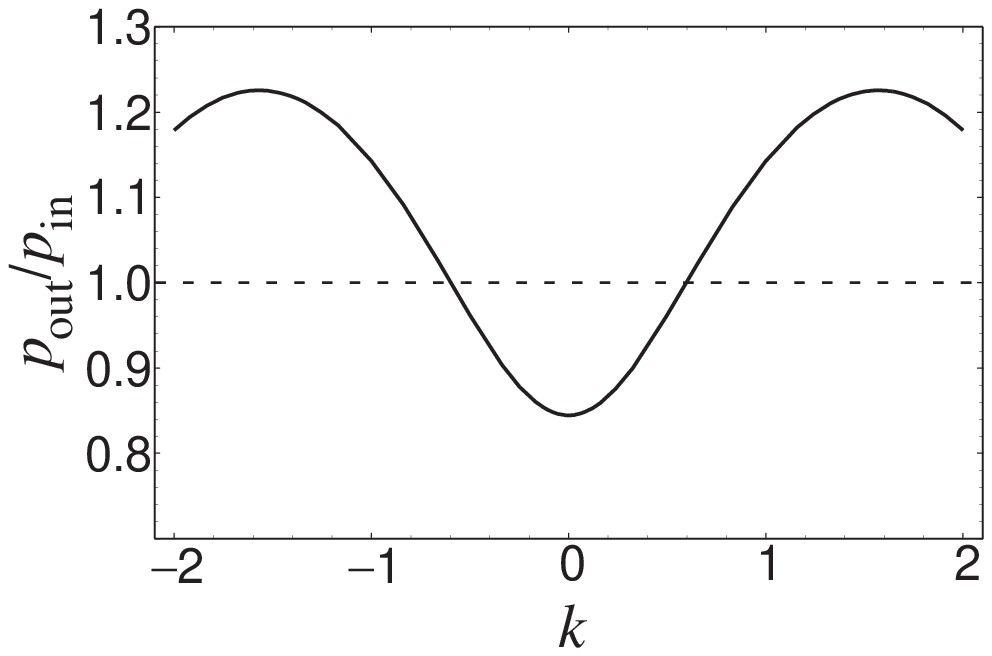}
  \caption{Purification efficiency for 
  the CSS with $\varphi=\pi$ conditioned on 
  the homodyne detection outcomes. The phase of the local oscillator 
  is chosen to measure the quadrature $k \equiv x_{\pi/2}$. 
  $T=0.5$, $\alpha=1$, and $p_{\rm in}=0.5$.} 
  \label{gainm}
 \end{center}
\end{figure}

\begin{figure}[h]
\begin{center}
  \includegraphics[width=0.9\linewidth]{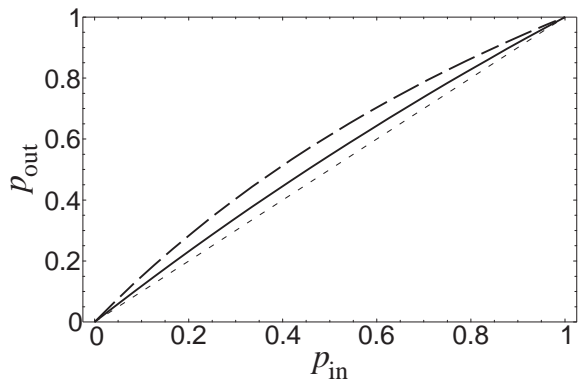}
  \caption{Dependence of the $p_{\rm out}$ on $p_{\rm in}$ for 
  the CSS with $\varphi=0$ (solid line) and $\varphi=\pi$ (dashed line) 
  where the outputs are conditioned on the measurement outcomes 
 $k=0$ and $k=\pi/2$, respectively. $T=0.5$ and $\alpha=1$. 
 The dotted line denotes $p_{\rm out}=p_{\rm in}$ as a reference. }
\label{gain-p}
 \end{center}
\end{figure}

Figs. \ref{fig:cond_k_p} and \ref{fig:cond_k_m} show these probability distributions for $\varphi=0$ and $\pi$, 
respectively. 
The probability distributions $P_0$, which is represented by the dashed curves, are Gaussian, while the probability distributions for the CSS, $P_{C_\varphi}$, (represented by solid lines) are non-Gaussian due to 
the quantum coherence between the two coherent state components. 
As discussed above, these distinct differences between the probability distributions 
enable the purification protocol: It succeeds when the absolute quadrature value falls within a certain threshold for $\varphi =0$ and outside a certain threshold for $\varphi =\pi/2$, since in these regions the probability for a CSS to occur is larger than the probability for $\hat{\rho}_0(\alpha)$ to occur.   
The efficiencies of the purification protocol for $\varphi=0$ and $\varphi=\pi$ 
are illustrated in Figs.~\ref{gainp} and \ref{gainm}, respectively. 
In case of $\varphi=0$, the purification condition of 
\eqr{eq:condition} is satisfied in the range given by
$|\cos\theta| > e^{-2R\alpha^2}$ and the efficiency of the purification protocol is optimised for $\theta=0$. 
The protocol works more efficiently when the initial CSS has the phase $\varphi=\pi$. In that case, 
purification is achieved in the range given by 
$|\cos\theta| < e^{-2R\alpha^2}$ and is maximized at $\theta=\pi$ 
where the initial input phase $\varphi=\pi$ is transformed into 
$\varphi=0$ after purification.
The maximum increase of $p_{\rm out}$ as a function of $p_{\rm in}$ for $T=1/2$ and $\alpha =1$ 
is depicted in Fig.~\ref{gain-p}, and we see that the improvement for these parameters is highest around $p_{\rm in}=1/2$.

\begin{figure}[h]
 \begin{center}
  \includegraphics[width=0.9\linewidth]{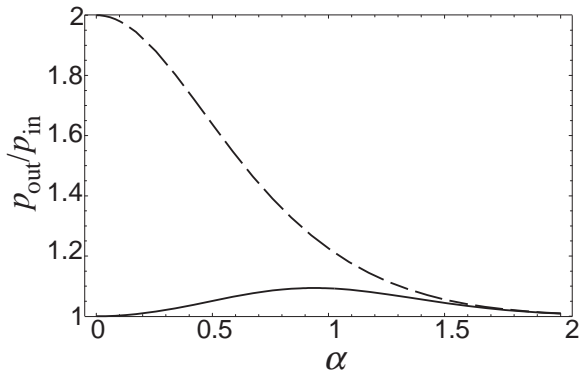}
  \caption{Dependence of the purification efficiency on the initial amplitude 
  for CSSs with $\varphi=0$ (solid line) and $\varphi=\pi$ (dashed line).
  $T=0.5$ and $p_{\rm in}=0.5$. The measurement outcome is selected to satisfy 
  the optimal condition $\varphi+\theta$=0.}
  \label{gain-alpha}
 \end{center}
\end{figure}

\begin{figure}[h]
 \begin{center}
  \includegraphics[width=1\linewidth]{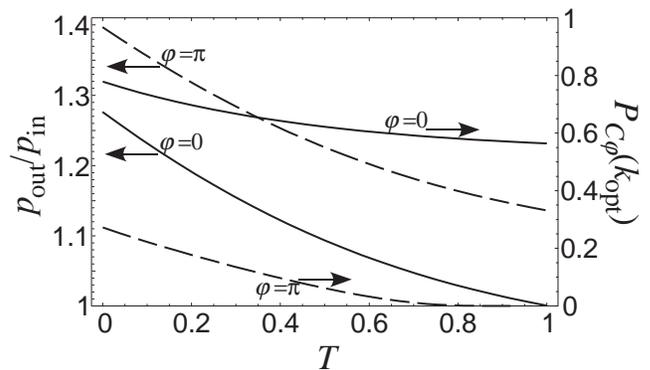}
  \caption{
  Dependences of the purification efficiency 
  $p_{\rm out}/p_{\rm in}$ (left) 
  and the probability density $P_{C_\varphi} (k_{\rm opt})$ 
  (right) on the transmittance of the tapping beam splitter $T$ 
  for $\varphi=0$ (solid line) and $\varphi=\pi$ (dashed line). 
  $p_{\rm in}=0.5$ and $\alpha =1$. The measurement outcome 
  $k_{\rm opt}$ is selected to satisfy the optimal condition 
  $\varphi+\theta=0$.}
  \label{gain-T}
 \end{center}
\end{figure}

In Figs.~\ref{gain-alpha} and \ref{gain-T}, we show 
how the purification efficiency depends on the amplitude $\alpha$ and 
the transmittance of the beam splitter $T$, respectively, 
for an input state with $p_{\rm in}=0.5$. 
From Fig.~\ref{gain-alpha}, we see that the purification of states with $\varphi =0$ (represented by the solid curve) is most effective around $\alpha \sim 1$ whereas for states with $\varphi =\pi$ the efficiency increases with smaller amplitudes. In Fig.~\ref{gain-T}, where the purification efficiency is plotted versus the transmission coefficient, it is evident that small transmission coefficients are advantageous. However, as already pointed out above, small transmission coefficients are on the other hand not desirable since it degrades the amplitude of the state.
Note that, obviously the purification does not work at $T=1$. 
Although, in Fig.~\ref{gain-T}, 
$p_{\rm out}/p_{\rm in}$ is still larger than 1 for $\varphi=\pi$ at $T=1$, 
the probability amplitude $P_{C_{\pi/2}} (0)$ goes to zero and thus 
this event never happens.  
These results, therefore, show that our scheme can be used to purify small amplitude CSSs, which are exactly the ones that recently has been produced \cite{wenger04}. 
However, the reduction of amplitude of the CSS is unavoidable 
in our scheme and it is therefore desirable to amplify the amplitude of the CSSs after purification. This is the topic of the next section.

Before ending this section, we address the effect of practical losses 
in homodyne detection. 
In practice, the efficiency of homodyne detection is less than unity 
due to the imperfect quantum efficiency of photodiodes, misalignment, 
mode mismatch, losses of optics etc. 
Such non-ideal homodyne detection can be modeled by 
a lossy quantum channel defined in Eq.~(\ref{eq1.1}) plus 
a perfect homodyne detector \cite{Leonhardt}. 
The effect of this imperfection in our scheme is estimated by 
replacing $\eta$ with $\eta (T + \eta_H R)$ in Eq.~(\ref{eq:p}) where 
$\eta_H$ is the effective efficiency of the homodyne detection. 
However, it should be noted that the recent quantum network 
experiments in CW domain at the wavelength of Si-photodiode 
can achieve sufficiently high homodyne effciency, e.g. $\eta_H > 0.98$ 
in \cite{Yonezawa04}. 
For example, with $\varphi=\pi$, $T=0.5$, $\alpha=1$, and $p_{\rm in}=0.5$, 
the purified outputs will be $p_{\rm out}=0.613$ and $0.609$ for 
$\eta_H=1$ and $0.98$, respectively. 
(Although the experiment in \cite{wenger04} was performed in pulsed regime, 
the feasibility of non-Gaussian experiments with CW sources has also been 
discussed in \cite{SasakiSuzuki05} for example.)


%%%%%%%%%%%%%%%%%%%%%%%%%%%%%%%%%%%%%%%%%%%%%%%%%%%%%%%%%%%%%%
\section{Amplification of coherent superposition states\label{sec:5}}
%%%%%%%%%%%%%%%%%%%%%%%%%%%%%%%%%%%%%%%%%%%%%%%%%%%%%%%%%%%%%%
Lund {\it et al.} have proposed a scheme to conditionally produce 
a large amplitude CSS from two small amplitude CSSs 
by using two 50/50 beam splitters, an ancillary coherent state and two 
on/off photon detectors \cite{lund04}, and later on the scalability of this scheme was studied in details~\cite{jeong05}. 
They pointed out that when one uses a weakly squeezed single photon state 
corresponding to a small CSS (which  
in practice inevitably consists of a mixture between the single photon state and a vacuum state) it is possible to reduce the squeezed vacuum component during the amplification process. 
In this section, we discuss the application of such a scheme 
to obtain a larger amplitude CSS from two decohered CSSs. 
We show that, the decoherence is, unfortunately, in most cases increased 
as a result of the amplification process.

\begin{figure}[h]
\begin{center}
 \includegraphics[width=0.9\linewidth]{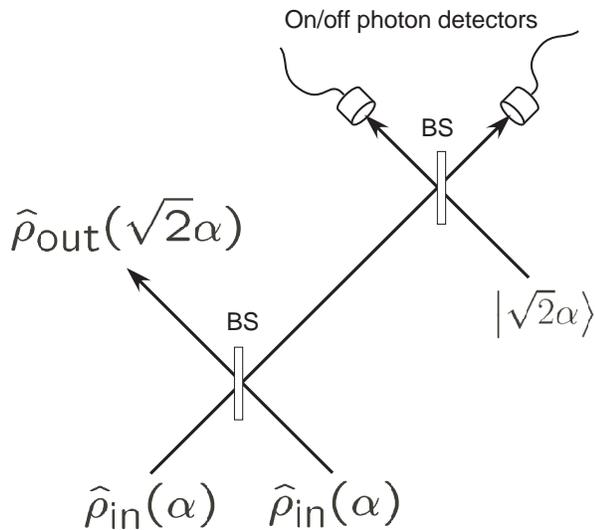}
 \caption{A schematic of the CSS-amplification process \cite{lund04} 
with the decohered input and output.}
 \label{fig:amplification}
\end{center}
\end{figure}

The scheme proposed in Ref.~\cite{lund04} is illustrated in
Fig.~\ref{fig:amplification}. 
Two copies of the CSSs with small amplitude are combined 
on a 50/50 beam splitter. 
Then one of the outputs is mixed with a coherent state with amplitude $\sqrt{2}\alpha$ on a 50/50 beam splitter. Subsequently the two outputs are measured using two on/off photon detectors. 
If both detectors click, the remaining quantum state 
is projected onto a larger CSS. In addition to the amplification of the amplitude, the projection process also filter out the squeezed vacuum component 
when the inputs are two mixed states of weakly squeezed single photon 
and vacuum.

Now, suppose that the input is two copies of 
the decohered CSS $\hat{\rho}_{\rm in} (\alpha)$ defined 
in \eqr{eq:initial_state}. Then, the input state is described by the tensor product
%%%%%%%%%%%%%%%%%%%%%%%%%%%%%%%%%%%%%%%%%%%%%%%%%%%%%%%%
\begin{eqnarray}
\label{eq:amp_input}
\hat{\rho}_{\rm in} \otimes \hat{\rho}_{\rm in} 
& = & p_{\rm in}^2 \hat{\rho}_{C_{\varphi}} \otimes \hat{\rho}_{C_{\varphi}} 
+ (1-p_{\rm in})^2 \hat{\rho}_0 \otimes \hat{\rho}_0
\nonumber\\ & & 
+ p_{\rm in} (1-p_{\rm in}) 
\left( \hat{\rho}_{C_{\varphi}} \otimes \hat{\rho}_0 + 
\hat{\rho}_{C_{\varphi}} \otimes \hat{\rho}_0 \right) ,
\nonumber\\
\end{eqnarray}
%%%%%%%%%%%%%%%%%%%%%%%%%%%%%%%%%%%%%%%%%%%%%%%%%%%%%%%%
and in the conditioned amplified output state, only the first term is mapped onto a large CSS whereas all other terms are 
transformed into a mixture of coherent states. 
As a result, the fraction of a CSS in the conditioned output 
states are approximately degraded as $p_{\rm out} \sim p_{\rm in}^2$. 
More precisely, the output probabilities $p_{\rm out}^+$ and $p_{\rm out}^-$ 
with inputs $\hat{\rho}_{C_0}(\alpha)$ and $\hat{\rho}_{C_\pi}(\alpha)$, 
respectively, are given by 
%%%%%%%%%%%%%%%%%%%%%%%%%%%%%%%%%%%%%%%%%%%%%%%%%%%%%%%%
\begin{equation}
\label{eq:amp_out}
p_{\rm out}^{\pm} = 
\frac{ \frac{1+e^{-4\alpha^2}}{(1 \pm e^{-2\alpha^2})^2} p_{\rm in}^2 }{
\frac{1+e^{-4\alpha^2}}{(1 \pm e^{-2\alpha^2})^2} p_{\rm in}^2 
+ \frac{ 2 p_{\rm in} (1-p_{\rm in}) }{ 1 \pm e^{-2\alpha^2} } 
+ (1-p_{\rm in})^2 }. 
\end{equation}
%%%%%%%%%%%%%%%%%%%%%%%%%%%%%%%%%%%%%%%%%%%%%%%%%%%%%%%%
Note that the conditioned output CSS with the probability 
$p_{\rm out}^{\pm}$ is always 
$|\psi_{C_{0}}(\sqrt{2}\alpha)\rangle$ \cite{lund04}. 
The parameter region where the inputs are 
purified during the conditional process can be evaluated from 
the condition $p_{\rm out}^{\pm}>p_{\rm in}$. First, we find that if the phase $\varphi$ is $0$ (that is $\hat{\rho}_{C_0}(\alpha)$ is amplified), the inequality cannot be satisfied. Therefore, in this case, the output probability $p_{\rm out}^+$ is always smaller than the input probability, or in other words, the decoherence is always amplified. However, if the amplitude of decohered CSS with $\varphi =\pi$ (that is $\hat{\rho}_{C_\pi}(\alpha)$) are amplified, then the inequality $p_{\rm out}^->p_{\rm in}$ can be satisfied if
%%%%%%%%%%%%%%%%%%%%%%%%%%%%%%%%%%%%%%%%%%%%%%%%%%%%%%%%
\begin{equation}
\label{eq:amp_purif_minus}
p_{\rm in} > \frac{1}{2} \left( e^{2\alpha^2} -1 \right)^2 . 
\end{equation}
%%%%%%%%%%%%%%%%%%%%%%%%%%%%%%%%%%%%%%%%%%%%%%%%%%%%%%%%
This implies that the average photon number of the input CSS must be 
$\alpha^2 < \ln (\sqrt{2}+1) /2 \approx 0.44$ for successful amplification of the amplitude without amplifying the decoherence.

\begin{figure}[h]
\begin{center}
 \includegraphics[width=1.0\linewidth]{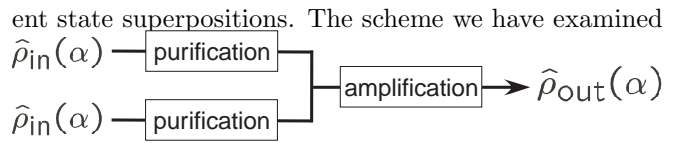}
 \caption{Concatenation of the purification and amplification processes.}
 \label{fig:concatenation}
\end{center}
\end{figure}

Finally, we investigate the possibility to repeat the purification process 
by concatenating this amplification scheme. 
The situation considered here is illustrated in Fig.~\ref{fig:concatenation}. 
Two initial states $\hat{\rho}_{\rm in}(\alpha)$ 
are purified in parallel by the purification processes with $T=1/2$. 
Once the purified outputs $\hat{\rho}_{\rm out} (\alpha/\sqrt{2})$ 
from both processes are simultaneously obtained,
the outputs are injected into the amplification process 
to obtain the final output which has the original amplitude $\alpha$. 
Let us now determine whether the output probability $p_{\rm out}$ 
is larger than the input probability $p_{\rm in}$ for such a machine:
Since the phase of the CSS obtained from the amplification process 
is always $\varphi = 0$, we assume the initial states to have $\varphi=0$. 
We then look for parameters where $p_{\rm out} > p_{\rm in}$ 
by replacing $p_{\rm in}$ in Eq.~(\ref{eq:amp_out}) by $p_{\rm in}/(p_{\rm in} + P_0/P_{C_\varphi} (1-p_{\rm in}))$ 
(which is the output probability from the purification step). However, unfortunately, a parameter interval fulfilling such criterion does not exist, and we conclude that for all possible amplitudes $\alpha$, the decoherence is amplified along with the amplitudes.
It suggests that when the goal is  
to generate large amplitude CSSs from smaller ones 
by using the conditional processes with linear optics proposed in Ref.~\cite{lund04}, 
we have to be careful about the effects of linear loss even if 
the initial states contain only a 
very small fractions of mixtures of coherent states. 
As we have shown, the decoherence is in most cases amplified during 
the conditional generation of large CSSs. 

%%%%%%%%%%%%%%%%%%%%%%%%%%%%%%%%%%%%%%%%%%%%%%%%%%%%%%%%
\section{Conclusions\label{sec:6}}
%%%%%%%%%%%%%%%%%%%%%%%%%%%%%%%%%%%%%%%%%%%%%%%%%%%%%%%%
In this paper, we have investigated a simple scheme to 
recover the quantum coherence of decohered coherent state superpositions. 
The scheme we have examined is based on partial estimation of the state by tapping off a part of it using a beam splitter, measuring the reflected part by homodyne detection and finally select out favourable events that maps the decohered state onto a purer state.

We derived the condition for the measurement to maximize 
the efficiency of the purification and showed that 
homodyne measurement is one of the optimal strategies. 
Since homodyne measurement is a well developed technique, 
our purification scheme is experimentally easier to implement than previously proposed purification protocols.
We therefore believe that our scheme can be directly implemented in current on-going experiments on the generation of CSS to compensate 
for losses and increase the nonclassicality of the states.

In addition, we showed that this simple measurement-induced 
scheme inevitably causes further amplitude degradation and 
is not suitable for the purification of large amplitude CSS 
with $\alpha \ge 2$. 
Moreover, we have discussed the possibility to restore the degraded 
amplitude by the linear optical scheme proposed in \cite{lund04}. 
If the initial CSSs are decohered due to a linear loss, 
the scheme is increasing not only the amplitude but also 
the degree of the mixture. 
We showed that, even combining this scheme with the purification process, 
it is still hardly possible to recover the coherence of CSS. 
Note that, also in the context of entanglement purification, 
it has been suggested in \cite{jeong02-qic} that 
the purification of entangled coherent states decohered 
by linear loss also seems to be difficult 
by using only linear optics and photon counting without ancillary CSSs. 
These results emphasize the importance of the further 
investigations of the purification scheme.

\begin{acknowledgements}
ULA acknowledges the financial support from the EU project COVAQIAL 
(FP6-511004).
\end{acknowledgements}

\end{document}